\documentclass[12pt,preprint]{aastex}
\usepackage{amsbsy}
\newcommand{\be}{\begin{equation}}
\newcommand{\ee}{\end{equation}}

\newcommand{\vn}{{\bf v}_{n}}
\newcommand{\vns}{{\bf v}_{ns}}
\newcommand{\vs}{{\bf v}_{s}}

\newcommand{\vv}{{\bf v}}

\newcommand{\pa}{\partial}
\newcommand{\om}{{\omega}}

\newcommand{\rna}{\rho_{n}}
\newcommand{\rsa}{\rho_{s}}

\newcommand{\Rey}{{\it Re}}
\def\ltsima{$\; \buildrel < \over \sim \;$}

\def\lsim{\lower.5ex\hbox{\ltsima}}
\def\gtsima{$\; \buildrel > \over \sim \;$}
\def\gsim{\lower.5ex\hbox{\gtsima}}

\begin{document}
\title{Global three-dimensional flow of a neutron superfluid  in a
spherical shell in a neutron star}

\author{C. Peralta,\altaffilmark{1,}\altaffilmark{2} A. Melatos,\altaffilmark{1} M. Giacobello\altaffilmark{3} and A. Ooi\altaffilmark{3} }

\email{cperalta@physics.unimelb.edu.au}

\altaffiltext{1}{School of Physics, University of Melbourne,
Parkville, VIC 3010, Australia}

\altaffiltext{2}{Departamento de F\'{\i}sica, Escuela de Ciencias,
Universidad de Oriente, Cuman\'a, Venezuela}

\altaffiltext{3}{Department of Mechanical Engineering, University
of Melbourne, Parkville, VIC 3010, Australia}


\begin{abstract}
\noindent We integrate for the first time the hydrodynamic
Hall-Vinen-Bekarevich-Khalatnikov
equations of motion of a $^{1}S_{0}$-paired neutron superfluid in a rotating
spherical shell, using a
pseudospectral collocation algorithm coupled with a time-split
fractional scheme. Numerical instabilities 
are smoothed by spectral filtering. Three numerical
experiments are conducted, with the following results. (i) When the
inner and outer spheres are put into steady differential rotation, the
viscous torque exerted on the spheres oscillates quasiperiodically
and persistently (after an initial transient). The fractional oscillation amplitude ($\sim 10^{-2}$)
increases with the angular shear and decreases with the gap width.
(ii) When the outer sphere is accelerated impulsively after an interval
of steady differential rotation, the torque increases suddenly, relaxes
exponentially, then oscillates persistently as in (i). The relaxation 
time-scale is determined principally by the angular velocity jump, 
whereas the oscillation amplitude
is determined principally by the gap width.
(iii) When the mutual friction force changes suddenly from
Hall-Vinen to Gorter-Mellink form, as happens when a rectilinear
array of quantized Feynman-Onsager vortices is destabilized by
a counterflow to form a reconnecting vortex tangle, the
relaxation time-scale is reduced by a factor of $\sim 3$ compared to (ii), and
the system reaches a stationary state where the torque oscillates with
fractional amplitude $\sim 10^{-3}$ about a constant mean value.
Preliminary scalings are computed for observable quantities like 
angular velocity and acceleration
as functions of Reynolds number, angular shear, and
gap width. The results are applied to the timing irregularities 
(e.g., glitches and timing noise) observed in radio pulsars.
\end{abstract}


\section{Introduction}
 The {\rm global} flow pattern in the superfluid interior of a neutron
star is poorly known, yet it is a  necessary input into {\rm
global} models of pulsar timing irregularities like glitches
\citep{sl96,lss00} and timing noise \citep{hobbs02}. The
importance of the global dynamics was first demonstrated in
pioneering laboratory experiments by \citet{tsatsa80}, who
impulsively accelerated rotating spherical and cylindrical
containers filled with He II and observed abrupt changes in
angular velocity, followed by intervals of prolonged relaxation,
reminiscent of glitches observed in rotation-powered pulsars.
The relaxation process can be interpreted as Ekman pumping
following abrupt spin up \citep{alpar78,aprs78,ae96}. The
laminar, linear spin up of He II between two parallel plates was
treated by \citet{r93}, who found a relation between the poloidal
secondary flow and the torque on the container.

In this paper, we seek to determine which, if any, of the basic
features of pulsar timing irregularities result directly from
the nonlinear hydrodynamics (e.g., oscillations and instabilities) of the 
global flow.
To do this, we solve numerically the
Hall-Vinen-Bekarevich-Khalatnikov (HVBK) equations \citep{hv56a,hv56b,bk61} 
for a
He-II-like superfluid contained in a differentially 
rotating spherical shell, building on previous numerical
simulations of viscous spherical Couette flow in nonastrophysical
settings \citep{mt87a,mt87b,dl94,mt95}. 

Spherical Couette flow is controlled by two parameters: the aspect ratio 
(dimensionless gap width) $\delta$,  and the Reynolds number $\Rey$. In
a slow rotator (${\it Re} \ll 1$), the flow is axisymmetric. In a fast
rotator ($\Rey \gg 1$), the flow is a combination of a primary
azimuthal rotation and a secondary meridional circulation induced
by Ekman pumping \citep{green68}. Transitions from the basic, null
vortex state to single and twin vortex states can be triggered by
gradually increasing $Re$ above a critical value, producing abrupt
changes in the rotation of the container. This has been observed
numerically and experimentally in small ($\delta < 0.24$) and
large ($\delta \geq 0.24$) gaps in ordinary fluids
\citep{ybm75,w76,ttuckerman, sn01}. \citet{h98} also found
evidence of time-dependent, asymmetric Taylor
vortices, in medium gaps ($\delta =0.336$). However, superfluid
analogs of these experiments are hard to perform, partly because
of problems with visualization techniques \citep{bs93}.
Numerical HVBK simulations have been performed in cylindrical
Taylor-Couette geometries \citep{hbj95,h01,hb00}, delivering good
agreement with experiments, but not yet in the more challenging
spherical Couette geometry. We attempt this here for the first time.

The paper is organized as follows. 
In \S\ref{sec:MODELO}, we establish the astrophysical context of
our numerical experiments by explaining how the idealized
spherical Couette problem relates to a realistic neutron star.
HVBK theory is
reviewed in \S\ref{sec:HVBKT}. 
The numerical method in spherical
geometry is described and verified against test cases in
\S\ref{sec:nummet}. Results are then presented for the long-term
response of the superfluid in three scenarios: the outer and inner
spheres rotate steadily and differentially (\S\ref{sec:steady}),
the outer sphere is accelerated impulsively (\S\ref{sec:impulse}), and
the mutual friction force changes
suddenly (from anisotropic to isotropic form) in response to a
line vortex instability (\S\ref{sec:instability}). Finally, in
\S\ref{sec:discussion}, we apply the  results to
pulsar timing irregularities and discuss the limitations of
our approach.

\section{Idealised hydrodynamic model of the outer core of a neutron star}
\label{sec:MODELO}
In this paper, we conduct numerical experiments that seek to mimic, in
an idealised context, the flow of neutron superfluid within the
{\it outer core} of a rotating neutron star. The density of the outer 
core lies in the range $0.6 \rho_* < \rho  < 1.5 \rho_*$,
where $\rho_* = 2.6 \times 10^{14}$ {\rm g cm}$^{-3}$ is the 
nuclear saturation density,
which translates to radii in the range 
$5 \, {\rm km} \lsim r \lsim 9 \, {\rm km}$ \citep{ss95,weber99,yls99,dhj03}.
We focus on the outer core because the superfluid is probably unpinned
in this region (see below) ---that is, it behaves essentially hydrodynamically---
and its physical state is reasonably well understood, unlike
the inner core. 
In the upper part of the outer core, $0.6 \rho_* < \rho < \rho_*$, the
neutrons form Cooper pairs in the $^1S_0$ state; in the
lower part,  $\rho_* < \rho < 1.5 \rho_*$, they pair in the  $^3P_2$ 
state \citep{weber99}.  
Here, for simplicity, we assume that the isotropic
$^1S_0$ phase fills the entire outer core.
The $^3P_2$ phase is anisotropic: gradients in
the orientation (texture) of the pair angular momenta
drive counterflows between the viscous and inviscid
components of the superfluid \citep{vollhardt,mm05}, introducing
complications (e.g. boundary conditions at
the $^1S_0$-$^3P_2$ interface) which lie outside the scope of this paper.
\footnote{A two-stream \citep{acp03} or 
Kelvin-Helmholtz \citep{mm05} instability can be excited
at the interface if the isotropic and anisotropic phases
are in relative motion.}

The outer core contains charged species, particularly
protons and electrons, which
are incorporated into the viscous component of the superfluid in 
our simulations.
The protons are probably in a type II superconducting state, but 
the electrons are not \citep{sauls89}.
In a type II superconductor, the protons inside the magnetic
fluxoids interact
with the Feynman-Onsager vortices in the
neutron superfluid, serving as natural pinning sites
\citep{sauls89};  the  pinning geometry can be complicated \citep{rzc98}
if the core magnetic field has comparable
poloidal and toroidal components \citep{td93}.
\footnote{Pinning in the outer core is arguably
inconsistent with the precession periods ($\sim 1$ {\rm yr}) observed in
some pulsars \citep{sls00}, because the Feynman-Onsager vortices
damp the precession on short time scales ($\sim 1$ {\rm hr})
as they pass through the array of magnetic fluxoids \citep{l03}.}
We neglect this effect in what follows, as well as the effects
arising from entropy gradients \citep{mm05}.
In addition, the protons exert an effective drag on the neutrons
by the so-called entrainment effect \citep{m91a,ss95}, in which the momentum
of one species is partly carried along by the other species,
as in $^3$He-$^4$He mixtures \citep{ab76,ac01}. We neglect
this effect in the computations in this paper, where there
is no entropy gradient at the $^3P_2$-$^1S_0$ interface \citep{mm05}.

The hydrodynamic boundary conditions at the upper and lower surfaces
of the outer core are set by the conditions in the inner crust
and inner core respectively. The physical state of the inner core
is essentially unknown, so we experiment with the two extremes of rough
walls (pinning) and perfect slip (no pinning) at this boundary. For
example, pinning is favored if 
triple-flavor color superconductivity results in a crystal lattice in the 
inner core \citep{r02}.
In the inner crust, it is widely believed that the Feynman-Onsager
vortices are trapped at nuclear pinning sites \citep{sauls89}, so
that a rough wall is the appropriate boundary condition. However,
recent work suggests that, even in the inner crust,
pinning may not occur. \citet{dp03}
studied the vortex-nucleus interaction 
and found intersticial pinning only, with low pinning forces
$\lsim 0.4$ {\rm MeV fm}$^{-1}$.

In the standard picture of pulsar glitches, the stellar crust is
loosely coupled to a rapidly rotating superfluid interior  \citep{lg98}.
Although we do not attempt to explicitly construct a model of glitches in
this paper, our simulations address a similar scenario: the
upper and lower surfaces of the outer core rotate
differentially, as the crust spins down electromagnetically, until
a glitch occurs and the crust suddenly accelerates.
We study the global superfluid flow pattern in the outer
core before and after the glitch. We also study how the Feynman-Onsager
vortices switch from a rectilinear array to a reconnecting tangle 
via the Donnelly-Glaberson instability \citep{gjo74}.
The transition from an organized to a disorganized distribution of 
superfluid vorticity in the outer core,
together with the acceleration of the crust, strongly
affects the rotation of the star, as we show below.

\section{HVBK theory}
\label{sec:HVBKT}
HVBK theory is a generalization of the two-fluid
Landau-Tisza theory for superfluid He II that includes the physics
of quantized Feynman-Onsager vortices \citep{hv56a,hv56b,bk61}. It applies equally well
to a neutron superfluid with $^1S_0$ parity \citep{tt86}. Fluid
particles in the theory are assumed to be threaded by many
coaligned vortices. In the continuum limit, we can define a
macroscopic superfluid vorticity $\om_s = \nabla \times
\vs$ which is not zero, despite the fact that, microscopically, the superfluid
obeys $\nabla \times \vs = 0$ . This is valid if the
length scales in the flow are longer than the average separation
between vortex lines. The isothermal HVBK equations of motion take
the form \citep{bj88,hb00}
\begin{equation}
\label{eq:supeq1} \frac{d_n \vn}{dt} = -\frac{\nabla p_n}{\rho}
+ \nu_{n} \nabla^{2} \vn + \frac{\rsa}{\rho} {\bf F} -
\frac{\rho_s \nu_s}{\rho} \nabla {|\om_s|},
\end{equation}

\begin{equation}
\label{eq:supeq2}  \frac{d_s \vs}{dt} = -\frac{\nabla p_s}{\rho}
+ \nu_{s} {\bf T} - \frac{\rna}{\rho} {\bf F} - \frac{\rho_s
\nu_s}{\rho} \nabla {|\om_s|},
\end{equation} with $d_{n,s} /dt =
\pa /{\pa t} + \vv_{n,s} \cdot \nabla$,
supplemented by the incompressibility condition $\nabla \cdot
\vn = \nabla \cdot \vs =0$, a good approximation in a neutron star,
where the flow is subsonic. In (\ref{eq:supeq1}) and
(\ref{eq:supeq2}), $\vv_{n,s}$ and $\rho_{n,s}$ are the normal
fluid and superfluid velocities and densities respectively. 
Effective pressures
$p_{s}$ and $p_{n}$ are defined by $\nabla p_s = \nabla p -
\frac{1}{2} \rho_n \nabla (\vns^2)$ and $\nabla p_n = \nabla p +
\frac{1}{2} \rho_s \nabla (\vns^2)$, with $\vns = \vn-\vs$.
We define $\nu_n$ to be  the kinematic viscosity of the normal fluid and
$\nu_{s} = (\kappa/4\pi) \log(b_0/a_0)$ to be the stiffness
parameter, where $\kappa = h/2 m_n$ is the quantum of circulation,
$m_n$ is the mass of the neutron, $a_0$ is the radius of the
vortex core, and $b_0$ is the intervortex spacing. Note that
$\nu_s$ has the same dimensions as a kinematic viscosity but
controls the oscillation frequency of Kelvin waves excited on the
vortex lines \citep{hbj95}.

The vortex tension force per unit mass, $\nu_{s} {\bf T}$, is
given by \begin{equation} \label{4h} {\bf T} = \om_{s} \times (\nabla \times
\hat{\om}_{s}), \end{equation} with $\hat{\om}_{s} = \om_{s}/|\om_{s}|$. The
tension arises from local circulation around quantized vortex lines. It
corresponds to the Magnus self-force exerted on a line by its
own, self-induced velocity field, which is given by $\nu_s \om_{s}
\times (\nabla \times \hat{\om}_{s})$ in the local induction
approximation of the full Biot-Savart integral, when the radius of
curvature is much larger than $a_0$ \citep{donnelly91}.

The mutual friction force per unit mass arises from the
interaction between the quantized vortex lines and the normal
fluid (via roton scattering in He II and electron scattering in a
neutron star) \citep{hv56a,hv56b}. The form of this force depends
on the global configuration of the vortices. If the
vortices ccupy a rectilinear array, the friction force per unit
mass takes the Hall-Vinen (HV) form \citep{hv56a,hv56b}

\begin{equation} \label{eq:ffhv}
 {\bf F} = \frac{1}{2} B \hat{\om}_{s} \times (
\om_{s} \times \vns - \nu_{s} {\bf T}) + \frac{1}{2}B^{\prime} (\om_{s}
\times \vns - \nu_{s} {\bf T})
\end{equation} where $B$ and $B^{\prime}$ are dimensionless, temperature-dependent 
coefficients \citep{bdv83}. However, in a realistic neutron star,
the distribution of vorticity  need not correspond to a
rectilinear vortex array \citep{g70,g74}. For example,
differential rotation can trigger the Donnelly-Glaberson (DG)
instability \citep{ccd73,gjo74}, in which a dense tangle of
vortices forms in the counterflow. Experimental evidence for
this effect comes from counterflow experiments that measure the
attenuation of second sound in narrow channels \citep{vinen57c,
sbd83}, supported by numerical simulations  using the vortex filament method
\citep{tab03}. From these experiments, it is inferred that the
friction force per unit mass takes the isotropic Gorter-Mellink
(GM) form \citep{gm49}

\begin{equation} \label{eq:gmforce} {\bf F} = A^{\prime} \left( \frac{\rho_{n} 
\rho_{s} v_{ns}^{2}}{ \kappa \rho^{2}}\right) \vns, \end{equation} where 
$A^{\prime} = B^3 \rho_n^2 \pi^2 \chi_1^2/3\rho^2 \chi_2^2$ 
is a dimensionless, temperature-dependent coefficient, related
to the original GM constant (usually denoted as
$A$ in the literature) by $A^{\prime} = A \rho \kappa$. Here, 
$\chi_1$ and $\chi_2$ are dimensionless constants of order unity 
\citep{vinen57c}.
Note that (\ref{eq:supeq1}) and (\ref{eq:supeq2}), derived
assuming a dense, roughly parallel array of vortices, may not be
consistent with (\ref{eq:gmforce}). The validity of HVBK theory in
turbulent flow is yet to be established \citep{bsd95}.

\section{Numerical Method}
\label{sec:nummet}

\subsection{Pseudospectral collocation in spherical geometry}
Equations (\ref{eq:supeq1}) and (\ref{eq:supeq2}) are discretized
in space using a pseudospectral collocation method
\citep{canuto88}. A time-split fractional step  algorithm
advances the solution \citep{canuto88,bc89,sh91}.
Nonlinear terms are treated explicitly using a third-order
Adams-Bashforth scheme and diffusive terms are treated implicitly
using a Crank-Nicholson scheme \citep{boyd02}. Incompressibility
is enforced by pressure-correction projection
 \citep{bcm01,chorin68}. We follow closely the approach
of \citet{bb02}, expanding the polar
($\theta$) and azimuthal ($\phi$) coordinates in Fourier functions 
and the radial ($r$) coordinate in Chebyshev polynomials. The velocity
fields are expressed as $u (r,\theta,\phi) = 
\sum_{i,j,k}
C_{ijk} T_{i}(r) f(j \theta) e^{i k\phi}$, where $T_i(r)$ is the
$i$-{\rm th} Chebyshev polynomial, $f(\theta)$ equals  $\sin \theta$ or 
$\cos\theta $, 
and $i, \,j, \,k$ are integers running from $0$ to 
$N_r, N_{\theta}$, $N_{\phi}$ respectively. The pole parity
problem which plagues Fourier series on a sphere \citep{boyd02}
must be handled with care; the tangential and azimuthal components
of $\vn$ and $\vs$ change sign across the poles \citep{orszag74,
fornberg98}.

Spectral methods are global and therefore sensitive to boundary
conditions. The normal fluid $\vn$ satisfies no-slip and
no-penetration boundary conditions at the inner and outer surfaces
of the container. There is no general agreement in the literature
on what boundary conditions are suitable for $\vs$. \citet{k65}
suggested that vortex lines can either slide along, or pin to, the boundaries.
The tangential velocity $\vv_L$ of a vortex line relative to a
rough boundary is given by \citep{hr77}

\begin{equation} \label{eq:bcs1} ({\bf n} \cdot \om_s) \om_s \times \vv_{L}= 0,
 \end{equation} where ${\bf n}$ is the unit normal to the surface; in this
case, vortex lines are permanently attached to the surface. At
a smooth boundary, on the other hand, one has

\begin{equation} \label{eq:bcs2}({\bf n} \cdot
\om_s) \om_s \times {\bf n}= 0; \end{equation} in this case, vortex lines are  oriented perpendicular
to the surface. Condition (\ref{eq:bcs1}) is difficult to
implement in HVBK theory, in which each fluid element is threaded by
many vortex lines, because $\vv_L$ cannot be calculated from
$\vv_{n}$ and $\vv_{s}$. On the other hand, frictionless sliding 
($\om_{s} \times {\bf n} = 0$) leads to numerical instabilities,
because $\om_s$ can develop parallel components near the
surface as it evolves \citep{hb00}. Therefore, in this paper, we adopt no-slip 
and no-penetration [i.e., pinning; see \citet{bel92}] boundary conditions 
for $\vs$ to stabilize the evolution.

Spectral methods can develop global oscillations, due to the 
Gibbs phenomenon, when
applied to problems with stiff numerical solutions or discontinuities \citep{vandeven91}. 
In the superfluid, such oscillations are damped indirectly (and
weakly) through ${\bf F}$.
To mitigate this problem, we filter out
high spatial frequency modes in the $r$ and $\theta$ expansions by
multiplying the spectral coefficients by an exponential filter defined by
$\sigma_e(k/N_{r,\theta})  =\exp[- 
(k/N_{r,\theta})^{\gamma} \ln \epsilon ] $, with $0
\leq |k| \leq N_{r,\theta}$, where 
$\epsilon =  2.2 \times 10^{-16}$ is the machine zero 
and $\gamma$ is the order of the filter.
Strong filtering is needed ($\gamma \leq 8$) in order to
stabilize the evolution. Most of our runs employ grids with
$(N_r,N_{\theta}, N_{\phi}) = (210, 250, 4)$, although a few have
$N_{\phi} = 32$.

As our simulations are the first to treat the spherical Couette problem
for superfluids, we are obliged to validate the code in the simpler case
of a classical viscous fluid. 
In this limit ($\rho_s \rightarrow
0$, $\rho_n \rightarrow \rho$ as $T \rightarrow T_c$), we succesfully 
reproduce the
$0 \rightarrow 1$ and $1 \rightarrow 2$ vortex-state transitions observed in
small  and large gaps in spherical Couette flow 
\citep{mt87a,mt87b,dl94}, achieving agreement to three significant
digits in the critical value of $\Rey$ and nine
significant digits in the torque.

\subsection{Model pulsar}
Our model pulsar consists of a spherical shell filled with a
$^1S_0$-paired neutron superfluid. The inner (radius $R_1$) and
outer (radius $R_2$) surfaces of the shell rotate at angular
velocities $\Omega_1$ and $\Omega_2$, respectively, about a 
common axis. Henceforth,
all quantities are expressed in dimensionless form using $R_2$
as the unit of length and $\Omega_1^{-1}$ as the unit of time. The 
Reynolds number and dimensionless gap width are defined
by $Re = \Omega_1 R_2^2/\nu_n$ and $\delta = (R_2 - R_1)/R_2$ respectively.

Our choice of parameters is limited by computational
capacity. For example, in a realistic pulsar, one might expect
$10^{-9} \lsim \Delta \Omega / \Omega \lsim 10^{-6}$ in the lead-up 
to a glitch \citep{hthesis02},
with $\Delta \Omega = |\Omega_2-\Omega_1|$.
We take $0.1 \leq \Delta \Omega \leq 0.3$ instead, in order to 
observe the long-term effect of differential rotation (i.e., several full
``wraps" of the inner sphere relative to the outer sphere) over a 
time interval $0 \leq t \leq 90$ that is computationally
achievable. 

Initially, both fluids must satisfy the
incompressibility condition, with $(\vn)_{r} = (\vs)_r = 0$ 
at $ r = R_1$, $R_2$. We
choose the Stokes solution \citep{landau_fluidos} as the initial
condition, with $\vs = \vn$. We also take $\rho_s/\rho = 0.908$ 
and $\rho_n/\rho = 0.092$, corresponding to the superfluid and normal
fluid fractions in He II at a fiducial temperature $T/T_c = 0.66$
(in a realistic pulsar, one has $T/T_c \lsim 0.1$).
Importantly, the Reynolds number in a typical neutron star  can reach
$\Rey \sim 10^{11}$ \citep{mm05}, whereas, in
our simulations, we restrict ourselves to 
$10^{2} \leq \Rey \leq 10^{4}$ for two
reasons: for $\Rey \geq 10^{5}$, the normal fluid flow
becomes turbulent, and we cannot resolve it properly; and, when
$Re$ is smaller, a steady state is reached in a shorter time.

We set the stiffness parameter to $\nu_s = 10^{-5}$ {\rm cm}$^2$s$^{-1}$,
such that $\nu_s \ll \nu_n$ as in a young pulsar
\citep{g70}, even though a more realistic
value is $\nu_s = 10^{-3} $ {\rm cm}$^2$s$^{-1}$
[cf. $\nu_n \approx 10^2 \, {\rm cm}^{2} {\rm s}^{-1}$; \citet{mm05}].
In practice, $\nu_s$ cannot be chosen 
independently of $\delta$.
For a small gap ($\delta < 0.1$) or large $\nu_s$, the tension
force disrupts the flow and leads to numerical instabilities, in
which the stiff superfluid streamlines bend sharply at the walls.
This effect is suppressed in a cylindrical shell, where the
tension causes the superfluid to rotate as a column parallel
to the curved walls \citep{hb00}. Consequently, perfect slip boundary
conditions on $\vs$ may prove beneficial in the spherical problem; we
will investigate them in a future paper.

The HV friction force parameters $B$ and $B^{\prime}$ are unknown in
a neutron star. We choose $B = 1.35$ and $B^{\prime} = 0.38$, adopting
the He II values at $T/T_c = 0.66$ \citep{bdv83,donnelly91,db98}. 
The parameter $A^{\prime} = 5.8 \times 10^{-3}$ 
(with $\chi_1/\chi_2 = 0.3$)  at the same temperature
can be calculated from a fitting formula
derived by \cite{dthesis72}, which is consistent with previously
published 
experimental values \citep{vinen57c}. Stable
long-term evolution is difficult to achieve
for this value of $A^{\prime}$, so we take 
$A^{\prime} = 5.8 \times 10^{-2}$ instead. A more 
detailed study of
the dynamics for a wider range of $B$, $B^{\prime}$ and $A^{\prime}$ 
will be presented in a forthcoming paper.

\section{Steady differential rotation}
\label{sec:steady}
In this section, we describe the results of a control experiment
in which the the inner and outer
spheres rotate steadily  yet differentially, i.e., $\Omega_1$ and
$\Omega_2 \neq \Omega_1$
are held constant.
We consider four cases, identified by Au,
Bu, Cu, and Du in Table \ref{table:t1}, corresponding to medium and
large gaps ($\delta = 0.5,\, 0.3$) and small and medium shear
($ 0.7 \leq \Omega_2 \leq 0.9$). 
A few preliminary runs
with $N_{\phi} = 32$ confirm that the flow remains
axisymmetric, so we restrict the resolution to
$(N_r, N_\theta, N_\phi) = (150,400,4)$ in what follows.

\begin{figure*}
\plotone{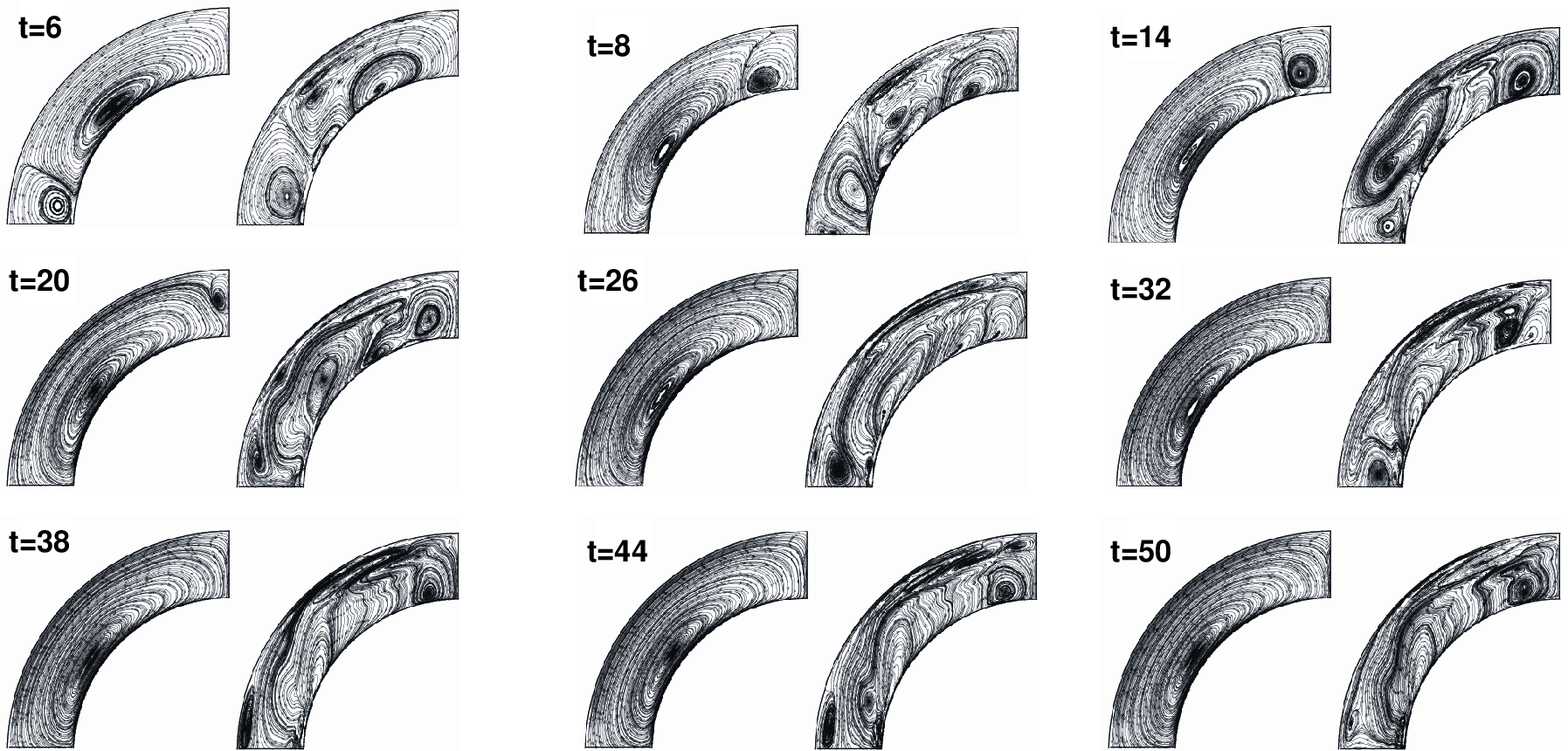} 
\caption{Meridional streamlines for the normal fluid (left)
and the superfluid (right) as a function of time $(6\leq t \leq 50)$ for $\delta=0.3$
and $\Delta \Omega = 0.3$ (case Bu, in Table \ref{table:t1}). 
The time is expressed in units of $\Omega_1^{-1}$.}
\label{fig:fig1}
\end{figure*}

\begin{figure*}
\plotone{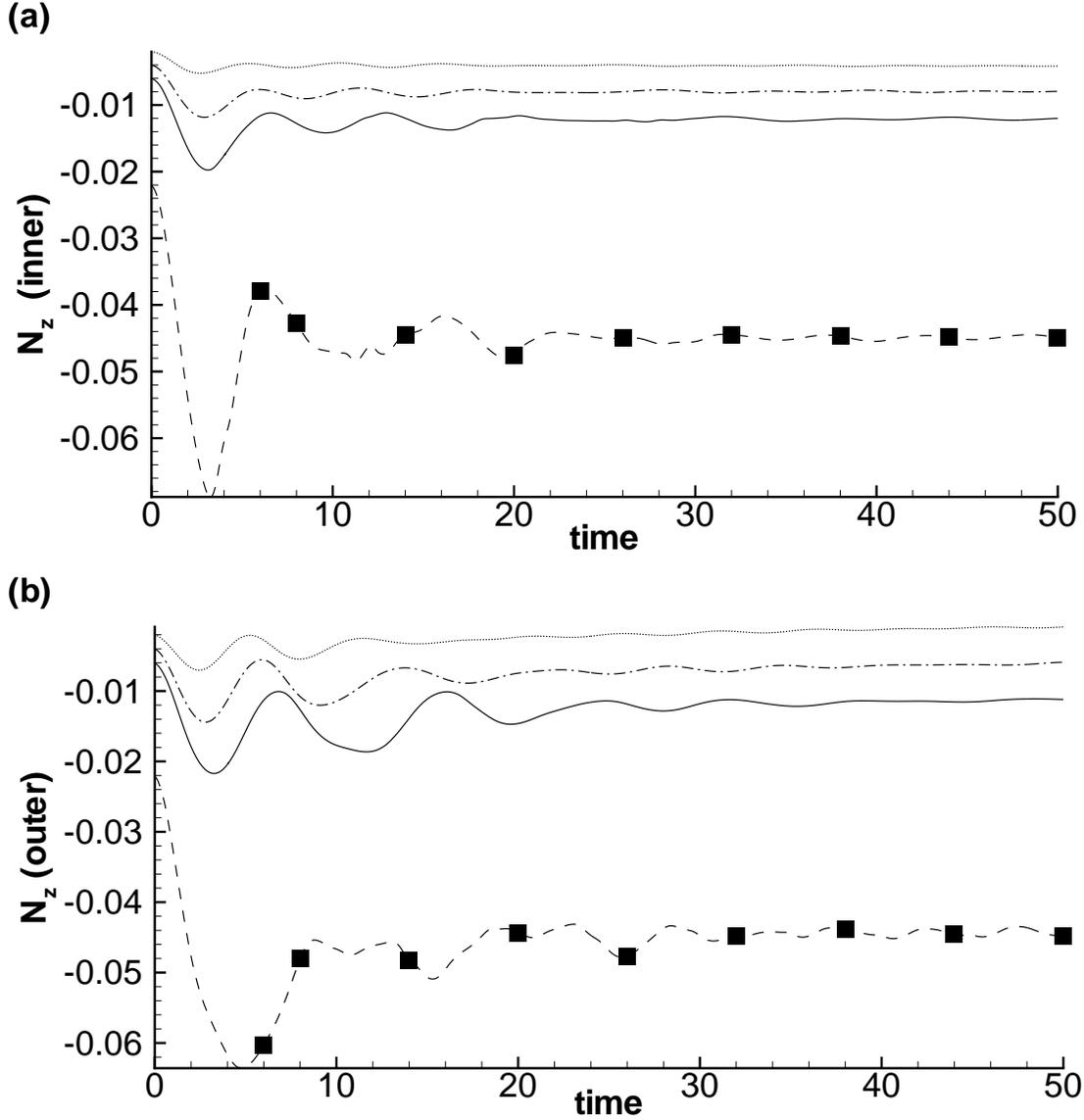} 
\caption{(a) Evolution of the (a)
inner torque and (b) outer  torque for the cases Au (solid curve),
Bu (dashed curve), Cu (dash-dotted curve), and Du (dotted curve), whose
parameters are quoted in Table \ref{table:t1}. The time is expressed in units 
of $\Omega_1^{-1}$ and the torques in units of $\rho R_2^5 \Omega_2^2$.
The filled squares correspond to the nine instantaneous snapshots of
the streamlines plotted in Figure \ref{fig:fig1}.} 
\label{fig:fig2}
\end{figure*}

Figure \ref{fig:fig1} displays the meridional streamlines 
of the normal fluid and the superfluid 
for $\delta=0.3$ and $\Delta \Omega = 0.3$ as a function of time.
The
corresponding torques on the inner and outer spheres are plotted 
as dashed curves in Figures \ref{fig:fig2}$a$ and \ref{fig:fig2}$b$
respectively, with squares 
marking the instants when the streamlines are plotted in Figure \ref{fig:fig1}.
The torque oscillates quasi-periodically, with a final
periodic state that persists as long as the differential rotation 
is maintained (up to $t=90$ in our runs), with a peak-to-peak fractional
amplitude $\Delta N_z/N_z \approx 0.02$. The flow pattern
in the normal fluid contains
a secondary vortex near the poles for $8 \leq t < 20$, which
expands and contracts quasi-periodically. For
$t > 20$, the secondary vortex disappears and a single meridional shell
remains, which also expands and contracts, but only slightly.
The superfluid 
streamlines
display a richer pattern: an equatorial vortex develops
at $ 6 \leq t \leq 8$, a polar vortex develops at $ 8 \leq t \leq 20$,
and the mid-latitude flow becomes disorganized for $t \geq 20$.
Note that we must reduce the spatial resolution 
to $(N_r, N_{\theta}) = (120,250)$ during the run
in order to stabilize the evolution; for
$(N_r, N_{\theta}) = (150, 400)$, one observes sudden
jumps in the torque unless $\Delta t \leq 10^{-5}$
(which lengthens the runs unacceptably).
It is possible that turbulence is generated in $\vs$ for $t \geq 20$ and
that, by intentionally choosing
a coarser resolution, we are effectively 
smoothing over the turbulent eddies.

The torque evolves similarly in all four cases. Initially, the linear
Stokes solution adjusts to a nonlinear, cellular solution (with
meridional circulation) within a time
$\Delta t \sim 6$. This transient state is of no interest
astrophysically. Subsequently, the torque oscillates persistently
---a new and astrophysically relevant phenomenon.
The oscillation amplitude depends strongly on the gap width $\delta$
and weakly on the shear $\Delta \Omega$. For example, if $\delta=0.5$
is held constant, the amplitude decreases from $\Delta N_z/N_z = 0.026$
to $\Delta N_z/N_z = 0.014$ as the shear decreases from $\Delta \Omega = 0.3$
(solid curve in Figure \ref{fig:fig2}) to $\Delta \Omega = 0.1$
(dotted curve in Figure \ref{fig:fig2}). On the other hand, if $\Delta \Omega = 0.3$ is
held constant, the amplitude decreases from $0.036$ to $0.026$ as
$\delta$ increases from $0.3$ to $0.5$ (dashed and solid curves 
in Figure \ref{fig:fig2}).

\begin{table}
\caption{ \label{table:t1} Simulation Parameters}
\begin{tabular}{cccc}
\tableline
\tableline
 $\delta$ & $\Omega_2$ &  Force type & Identifier\\ \tableline
  $0.5$ & $0.7$ &  HV & Au \\ 
  $0.5$ & $0.7 \rightarrow 1$ & HV & Aa  \\  
  $0.5$ & $0.7$ & HV $\rightarrow$ GM & Af \\ \tableline
  $0.3$ & $0.7$  &  HV & Bu \\ 
  $0.3$ & $0.7 \rightarrow 1$  &  HV & Ba \\
  $0.3$ & $0.7$  &  HV $\rightarrow$ GM &  Bf\\ \tableline
  $0.5$ & $0.8$  &  HV & Cu \\ 
 $0.5$ & $0.8 \rightarrow 1$  &  HV & Ca\\
 $0.5$ & $0.8$  &  HV $ \rightarrow$ GM & Cf  \\ \tableline
 $0.5$ & $0.9$  &  HV & Du  \\
 $0.5$ & $0.9 \rightarrow 1$  &  HV & Da  \\
 $0.5$ & $0.9$  &  HV $ \rightarrow$ GM & Df \\
\tableline
\end{tabular}
\end{table}

\section{Impulsive acceleration of the outer sphere}
\label{sec:impulse} 
\begin{figure*}
\epsscale{0.8}
\plotone{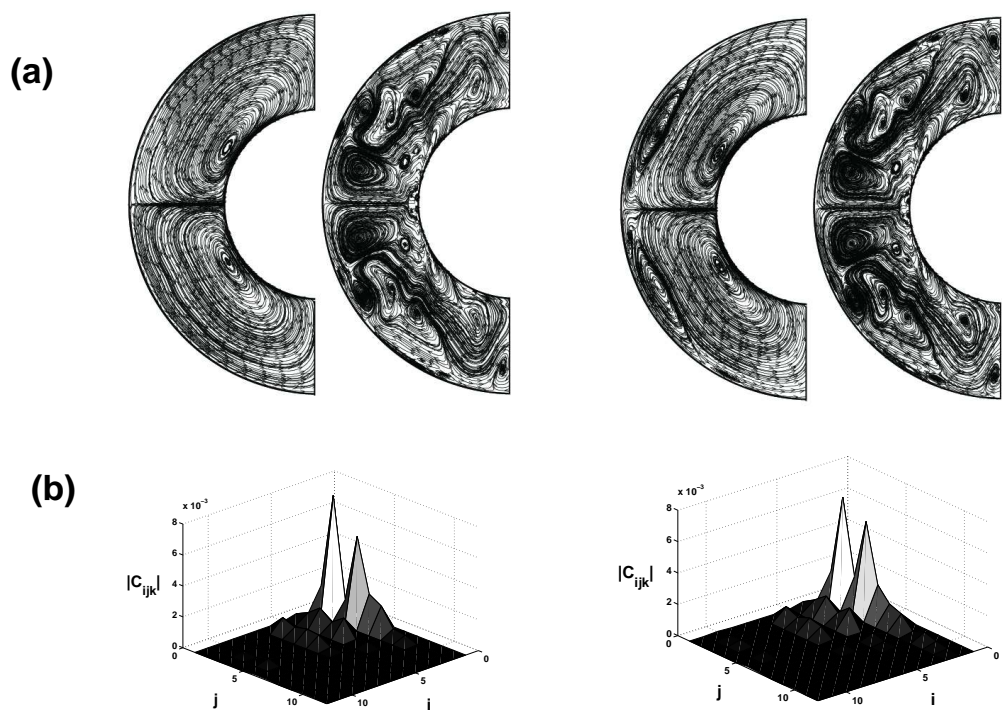}
\caption{(a) Meridional streamlines 
before (left pair of hemispheres) and after (right pair of hemispheres) 
the outer sphere is accelerated instantaneously from
$\Omega_2 = 0.7$ to $\Omega_2 = 1$ at $t=20$. For each pair of
hemispheres, the normal fluid streamlines are plotted at left and
the superfluid streamlines are plotted at right. The results
are for the case Af in Table \ref{table:t1}. (b) Histograms of the magnitudes
of the $12$ dominant spectral coefficients of $(\vv_n)_{\theta}$ 
before (left) and after (right) the acceleration at $t=20$, as a function 
of the Chebyshev polynomial index $i$
and the polar Fourier index $j$, with azimuthal index $k = 1$.}
\label{fig:fig3}
\end{figure*}

When a glitch
is triggered in a rotation-powered pulsar, the crust and
consequently the superfluid interior are spin up impulsively. We simulate 
a spin-up event of this kind by abruptly accelerating the rotation 
of the outer sphere from $ 0.7 \leq \Omega_2 \leq 0.9$
to $\Omega_2 = 1$ at $t = 20$.
Figure \ref{fig:fig3}$a$ shows the change in the flow pattern of
the normal fluid and superfluid before and after this sudden
acceleration. The difference is more pronounced
in the normal fluid; a secondary vortex appears at mid-latitudes
in each hemisphere, adjacent to the outer shell.
The  jump in torque comes almost entirely from these mid-latitude regions.

Figure \ref{fig:fig1}$b$ shows how the dominant
spectral coefficients  $C_{ijk}$ ($k=1$) of $(\vv_n)_\theta$ 
behave before and after the spin-up event.
The three modes of highest amplitude before 
acceleration are $C_{221} > C_{421} > C_{521}$.
After acceleration, the same ordering is observed, but with 
amplitudes $ \approx 12 \%$ lower, and the
fourth-strongest mode switches from $C_{621}$ to 
$C_{321}$. By contrast, in similar data
for $(\vv_s)_\theta$, no change is seen in the ordering
of the top $10$ spectral coefficients, and their amplitudes
change by less than $10 \%$.
The superfluid is therefore largely unaffected by the glitch, 
as one might anticipate by inspecting Figure \ref{fig:fig3};
most of the effect is transmitted to the normal fluid
by the viscous Ekman layer adjacent to the outer shell. 

\begin{figure*}
\epsscale{0.8}
\plotone{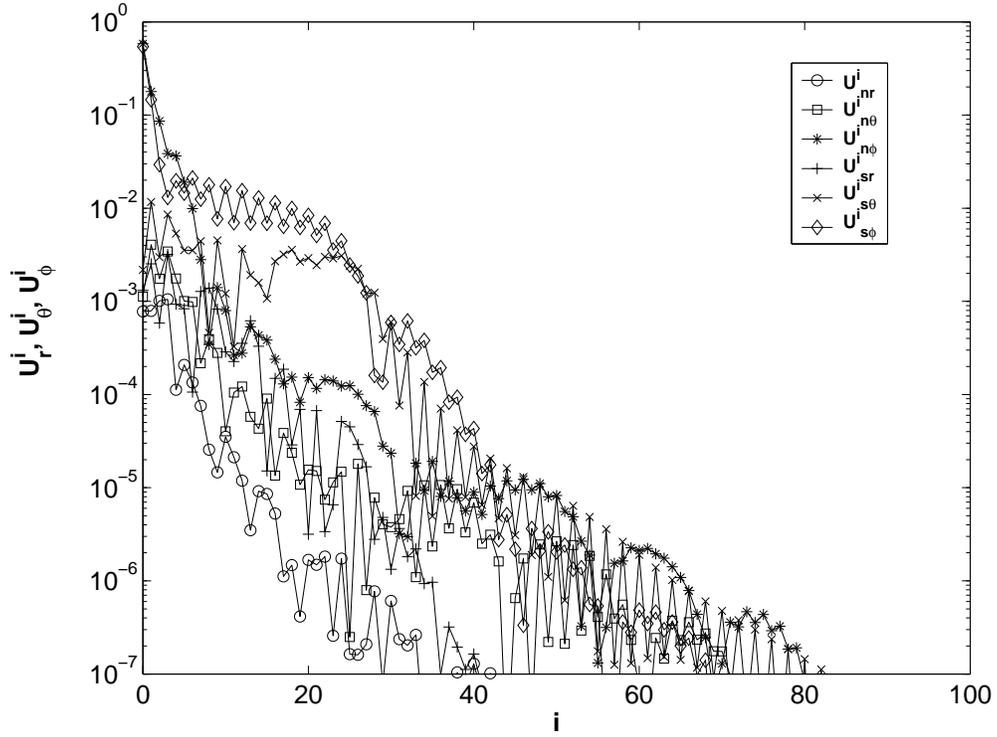}
\caption{Modes amplitudes for the $r$ ($\circ$), $\theta$ ($\square$), and
$\phi$ ($\ast$) components of the
normal fluid and the $r$ ($|$), $\theta$ ($\times$), 
and $\phi$ ($\diamondsuit$) components of the superfluid
as a function of the Chebyshev polynomial index $i$, at $t=20.1$. The
results are for case Af in Table \ref{table:t1}.}
\label{fig:fig4}
\end{figure*}

A key numerical issue is whether the spectral method
faithfully resolves the flow. Empirically, this occurs if
the mode amplitudes decrease quasi-monotonically
with polynomial index. We confirm this here.
The Chebyshev modes, plotted in Figure
\ref{fig:fig4}, decrease by a factor $\sim 10^{4}$ in amplitude from 
$i=1$ to $i=50$, indicating that the flow is resolved in $r$. 
We also increase
the resolution in $\phi$ at $t > 20$ by
Fourier interpolation and confirm that the results are identical 
for $4 \leq N_{\phi} \leq 32$.

When the outer sphere is accelerated, 
we find that the time-step must be halved to avoid sudden
artificial jumps in the torque. This property has been noted
widely in the literature: in viscous
spherical Taylor-Couette flow, the
nonlinear dynamics depends sensitively on the time-step
in the transition between different vortex states 
\citep{lk04,sn01,lbewr96}. For example,
\citet{lk04} reported that the transition to an asymmetric
$1$-vortex state can only be achieved by varying {\it Re} quasi-statically
near its critical value 
and using a small ($\Delta t = 10^{-4}$) time step. Different 
flow states in viscous spherical Couette flow are associated with
energy differences. If the system is driven through an
intermediate state too quickly, the flow cannot
follow the fast change, due to its inertia, and a steady final
state cannot be reached \citep{w76}.

\begin{figure*}
\plotone{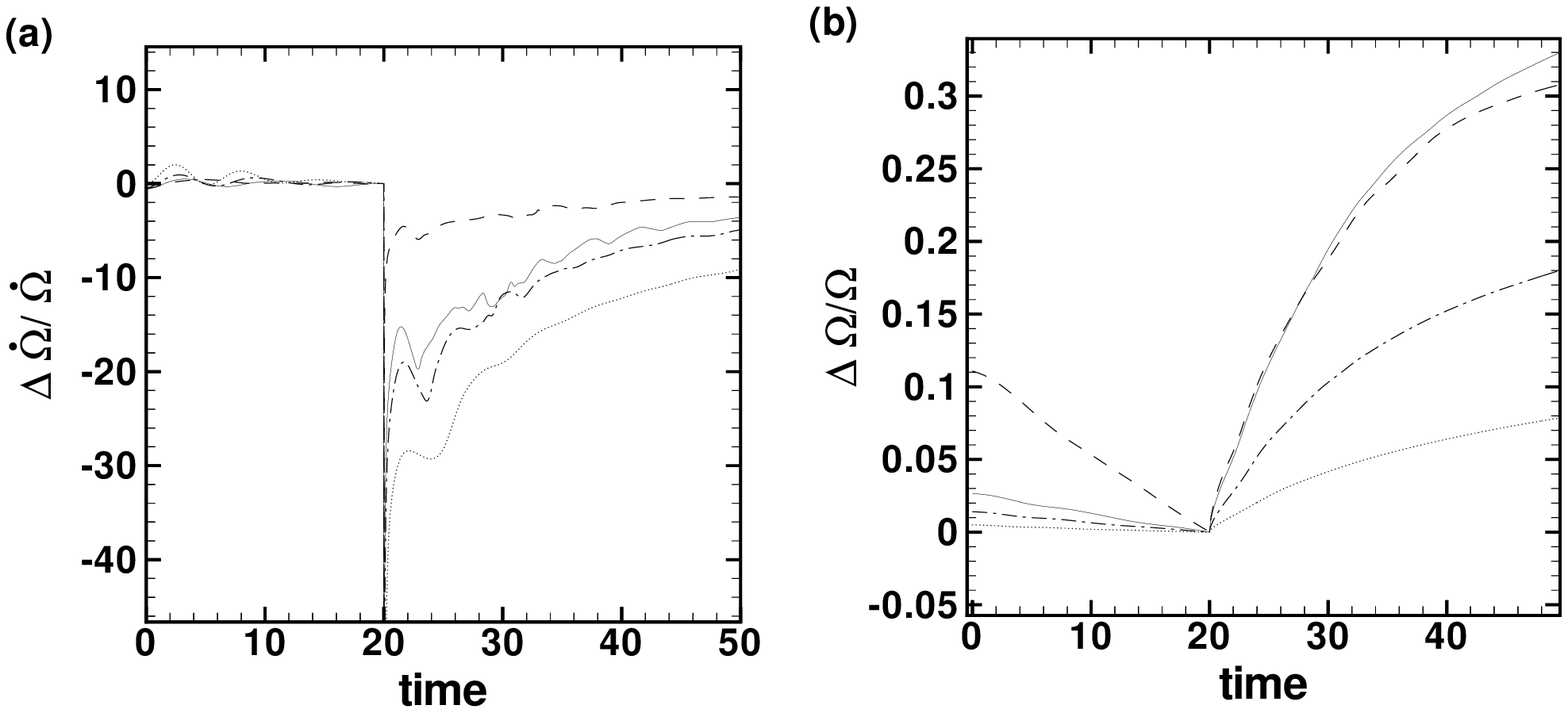} 
\caption{(a) Fractional change in angular acceleration of the
outer sphere, $\Delta \dot{\Omega}/\dot{\Omega} =
[\dot{\Omega}_2(t) - \dot{\Omega}_2(20)]/\dot{\Omega}_2(20)$,
as a function of time. (b) Fractional change in angular velocity
of the outer sphere,
$\Delta \Omega/\Omega = [{\Omega}_2(t) 
- {\Omega}_2(20)]/{\Omega}_2(20)$, as a function of time. Time is 
measured in units of $\Omega_1^{-1}$. Four cases are shown, whose
parameters are quoted in Table \ref{table:t1}: Aa (solid curve), Ba (dashed curve), Ca (dashed-dotted curve), and Da (dotted curve).}
\label{fig:fig5} 
\end{figure*}

The post-acceleration evolution of the torque 
is qualitatively similar in the four cases studied.
This can be observed in Figure \ref{fig:fig5},
where we plot the fractional change in the angular
velocity $\Delta \Omega / \Omega$ and its first time
derivative $\Delta \dot{\Omega} / \dot{\Omega}$ as a function of
time.\footnote{Note that $\Omega_{2} (t)$ is held fixed 
in the code after the acceleration at $t=20$. Hence the angular velocity
change $\Delta \Omega = \Omega_2(t) - \Omega_2(20)$ plotted
in Figure \ref{fig:fig5} is not computed self-consistently. It is 
found by integrating the viscous
torque on the outer sphere, calculated by solving (\ref{eq:supeq1})
and (\ref{eq:supeq2}) subject to the time-independent boundary
condition $\Omega_2(t)=$ {\rm constant}.} In radio pulsars, both 
these quantities are observed to
high accuracy in radio timing experiments.
The outer torque jumps at $t=20$, 
oscillates with a period
$ \Delta t \sim 5$, then decays quasi-exponentially.
Note that, in Figure \ref{fig:fig5}$b$, the curves for Aa, Ca and Da
are grouped together quite closely before acceleration, whereas
the curves for Aa and Ba are grouped closely after acceleration. This
implies that the gap width $\delta$ is the main parameter controling
the spin-down torque over the long term (i.e., during steady
differential rotation), as in \S\ref{sec:steady}, whereas the 
angular velocity jump $|\Omega_2 - \Omega_1|$
is the main parameter controling the relaxation time-scale.
By fitting an exponential to the secular evolution of  
$\Delta \dot{\Omega}$,
we find that the ($e^{-1}$) relaxation time-scale is given by
$\tau = 12$, $9.5$, $9.3$, and $8.2$ 
for Aa, Ba, Ca, and Da  respectively and scales roughly as 
$\tau \propto \delta^{1/3} |\Omega_2-\Omega_1|^{1/3}$, similar to
(but not exactly the same as) the Ekman scaling 
$\tau \propto |\Omega_2-\Omega_1|^{1/2}$.

\section{Instability of the Vortex Array}
\label{sec:instability} The stability of a rectilinear array
of quantized vortices
was investigated by Vinen in counterflow experiments in narrow
channels \citep{vinen57c}. He attributed the attenuation of
second sound to the generation of a ``vortex tangle"
in the counterflow. The critical (axial) counterflow velocity $v_{ns}$ for the 
onset of this turbulent state was calculated theoretically
by \citet{gjo74}, for
small perturbations of a single vortex line \citep{ccd73}. He found that, 
for $v_{ns} > 2 (2 \Omega \nu_s)^{1/2}$,
growing Kelvin waves are excited and the
vortex lines reconnect to form a tangle, in accord with experimental 
data \citep{ccd73}. 
Similar results have been produced by numerical simulations
with the vortex filament method \citep{s88, tab03}.
The inclusion of slow
rotation in counterflow experiments reduces the critical
$v_{ns}$ at which superfluid turbulence sets in \citep{sbd83}.

\begin{figure*}
\plotone{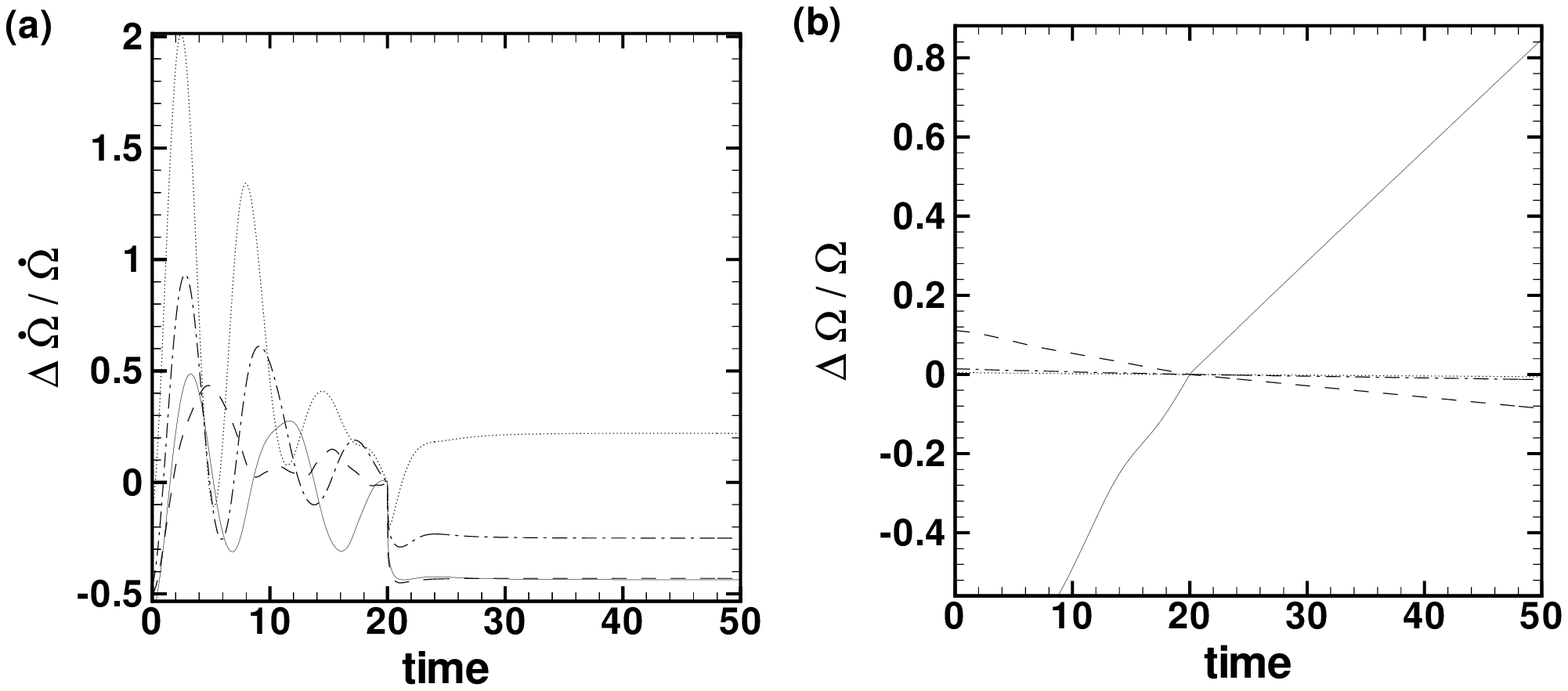} 
\caption{(a) Fractional change in angular acceleration of the
outer sphere, $\Delta \dot{\Omega}/\dot{\Omega} = 
[\dot{\Omega}_2(t) - \dot{\Omega}_2(20)]/\dot{\Omega}_2(20)$, as a function of time. (b) Fractional change in angular velocity
of the outer sphere, $\Delta \Omega/\Omega = [{\Omega}_2(t) 
- {\Omega}_2(20)]/{\Omega}_2(20)$, as a function of time.
Time is measured in units of $\Omega_1^{-1}$. Four cases are shown, whose
parameters are quoted in Table \ref{table:t1}: Af (solid curve), 
Bf (dashed curve), Cf (dashed-dotted curve), and 
Df (dotted curve).}
\label{fig:fig6} 
\end{figure*}

In order to simulate what happens when a vortex tangle develops from 
an initially
regular vorticity distribution, we suddenly change the friction force
in (\ref{eq:supeq1}) and (\ref{eq:supeq2}) from HV
to GM form, according to the prescription in \S\ref{sec:nummet}.
From Figure \ref{fig:fig6}, we observe that the torque reaches a stationary state $\sim 3$ 
times more quickly
than in the spin-up experiments
presented in \S\ref{sec:impulse}. The torque decays more quickly than
an exponential, as can be appreciated from Figure \ref{fig:fig6}.
Nevertheless, we can estimate the ($e^{-1}$) relaxation time-scale 
for the cases Af, Bf, Cf, and Df 
in Table \ref{table:t1} to be $3.4$, $3.4$, $3.0$, and 
$3.6$, respectively. The relaxation time is essentially independent
of $\delta$ and $|\Omega_2-\Omega_1|$.

Relaxation occurs more quickly than in the spin-up experiment
because the GM force is small compared to the 
HV force. From \S\ref{sec:impulse}, we have 
$|{\bf F}_{\rm GM}|/|{\bf F}_{\rm HV}| = (A^{\prime}/B)(\rho_n \rho_s/\rho^2)
(v_{ns}^2/\kappa \omega_s)$ [and hence $\propto (\rho_n/\rho)^{3}$ 
for $\rho_n \ll \rho$];
on the other hand, we find 
$|{\bf F}_{\rm GM}|/|{\bf F}_{\rm HV}| \approx 10^{-5}$ empirically
from our runs, implying $v_{ns}^2/\kappa \omega_s \approx 2.2 \times 10^{-6}$.
The sudden change
in the mutual friction effectively
uncouples the normal fluid and the superfluid,
and the torque evolves rapidly to a constant value
on a time-scale $ \Delta t \sim 3$.
After this time, the inner
and outer torques are equal to one part in $\sim 10^{3}$ and
their mean values hardly vary at all. Nevertheless, while
the torque appears constant for $t > 25$ in Figure \ref{fig:fig6}, 
closer examination on a magnified scale shows that $|N_2-N_1|$
oscillates persistently, out to $t=90$, with peak-to-peak amplitude $\Delta N_z/N_z \sim 10^{-3}$.

\section{Discussion}
\label{sec:discussion}

In this paper, we present the first three-dimensional
hydrodynamic HVBK simulations of a $^1S_0$-paired neutron superfluid 
in a rotating spherical shell, generalizing studies of
viscous fluids in spherical
Couette geometry \citep{mt87a,mt87b,dl94} and He II in
cylindrical Couette geometry \citep{hbj95,hb00}. The code 
accurately resolves the
superfluid flow pattern for Reynolds numbers and gap widths
in the range $10 \leq \Rey \leq 10^4$ and $0.2 \leq \delta \leq 0.7$,
respectively.

Persistent quasiperiodic oscillations are always observed in the
torque during steady differential rotation, after the initial
transient dies away, with typical period $\sim \Omega_1^{-1}$ and
fractional amplitude $\sim 10^{-2}$. The oscillation amplitude 
increases as $\Rey$ increases.
The gap width $\delta$ principally controls
the spin-down torque over the long term, during steady differential rotation, 
whereas the frequency jump $|\Omega_2 - \Omega_1|$ principally
sets the relaxation time-scale after a sudden acceleration of the
outer sphere.
If, instead of spinning up  the outer sphere, we suddenly switch the mutual 
friction force from anistropic (HV) form to
isotropic (GM) form, as happens when a rectilinear array of
quantized vortices reconnects unstably to form a tangle,
the relaxation 
time-scale is $\sim 3$ times shorter and
the torque difference $|N_2-N_1|$ oscillates subsequently with peak-to-peak
fractional amplitude $\sim 10^{-3}$ around a constant mean value.
We speculate that a 
transition from a turbulent flow, driven by long-term
differential rotation, to a laminar flow immediately after a glitch,
might cause some of the timing
irregularities observed in pulsars. Similar transitions from
turbulent to laminar flow in a superfluid have already been
observed in laboratory  experiments where He II, cooled to a few mK, flows
around an oscillating microsphere \citep{nks02,s04}. 

Computational limitations impose upon us some artificial approximations.
Some of these approximations will be tested independently in a future
paper. For example, one might hope to use a weaker filter when damping
the high spatial frequency modes, perhaps by introducing an artificial
viscosity \citep{alw94,ok96}, and one can model
small-scale turbulence using the techniques of large eddy
simulations \citep{gpmc91,l92,mcm04}. Moreover, although some of
our parameters are not as close as one might which to realistic
neutron star values, we are careful to respect the ordering of
physical quantities encountered in a neutron star (e.g., 
$\Delta \Omega_2 \ll \Omega_2$, $\Rey \gg 1$, $\nu_s \ll \nu_n$) and to
use values of the dimensionless friction coefficients ($B$, $B^{\prime}$
and $A^{\prime}$) that apply to the neutron star regime $\rho_n \ll \rho_s$
and are consistent with experimental data on He II.
By artificially reducing $\nu_s$ by two orders of magnitude,
we trade off some accuracy for computational speed in our modeling
of old pulsars, where the tension force is
the primary interaction \citep{g70}, although our modeling of
young pulsars is mostly unimpaired \citep{catalog04,lss00}.

We consider superfluid flow in the outer core of the star, where
there is no pinning in the body of the fluid, only at the boundary.
Consequently, those parts of our analysis pertaining to the formation
of a reconnecting vortex tangle via the DG instability may not translate
straightforwardly to the inner crust, where pinning at multiple
lattice sites along a vortex line may suppress the DG instability
and even prevent a tangle from forming at all. On the other hand,
if Ekman pumping is the primary core-crust coupling mechanism in
a real neutron star, then the dynamics in the outer core must dominate
the physics; crust-only mechanisms are inconsistent with
observations of Vela's glitches in the Ekman regime 
\citep{mhmk90,aeo96}.

The time-scales of torque fluctuations in our numerical 
experiments are generally 
$\lsim 10 \Omega_1^{-1}$, much shorter than the time-scales
for post-glitch relaxation and timing noise oscillations observed
in radio pulsars. However, it
is crucial to realize that the numerical time-scales will lengthen greatly
in more realistic simulations, where the angular shear is lower
and $\Rey$ is higher. Such numerical
experiments are computationally challenging, and we are investigating
alternative numerical approaches that may render them tractable.

A necessary extension of the code is to include radial
stratification. It has been argued that,
for a viscous fluid, stratification hampers the formation 
of an Ekman layer near the outer boundary
\citep{ae96}. The Brunt-V\"ais\"ala frequency in a neutron
star was estimated to be $ \omega_{\rm BV} \sim 500 {\rm s}^{-1}$
around nuclear density and $  \omega_{\rm BV} \sim0.7 \Omega_2$ 
for the Vela pulsar
\citep{rg92}, whereas one needs 
$ \omega_{\rm BV} \lsim 0.2 \Omega_2$ for
Ekman pumping to proceed effectively \citep{ae96}.
Stratification can also
affect the coupling between the crust and
the superfluid core of a neutron star by modifying precessional
modes \citep{ld04}. The pressure projection operation
in our code relies on incompressibility, so stratification
is hard to include self-consistently. As a first approximation, however,
one can set $v_r = 0$ in the incompressible calculation, as proposed 
by \citet{ld04}.

The HVBK equations may not apply to a vortex
tangle. An alternative theory which explicitly treats
displacements and oscillations of the vortex array was
developed by \citet{cb83,cb86}. The complexities thereby
introduced, and fresh uncertaintities over
boundary conditions, raise new numerical challenges.

\acknowledgments We acknowledge the computer time supplied by the
Australian Partnership for Advanced Computation (APAC) and the
Victorian Partnership for Advanced Computation (VPAC). We also 
thank George Hobbs for helpful discussions and
Caroline Andrzejewski for undertaking a thorough
literature review of spherical Couette flow in support of this research.


\newpage

\clearpage
 
\clearpage

\end{document}